\documentclass{emulateapj}
\usepackage{times}

\newcommand{\gray}{$\gamma$-ray}

\newcommand{\icrc}{Int.\ Cosmic Ray Conf.\ }

\newcommand{\app}{APh}
\newcommand{\sci}{{Science}}

\newcommand{\nimA}{Nucl.\ Instr.\ Meth.\ Phys.\ Res.\ A}

\newcommand{\plb}{Phys.\ Lett.\ B}
\newcommand{\jgpr}{J.\ Geo.\ Phys.\ Res.} 

\newcommand{\pubjournal}[5]{#1 #5, #2, #3, #4}
\newcommand{\pubjournala}[5]{#1 #5, #2, #4}

\newcommand{\geant}{GEANT4}
\newcommand{\ieeenuc}{IEEE\ Trans.\ Nuc.\ Sci.}

\shorttitle{The Gamma-Ray Albedo of the Moon}
\shortauthors{Moskalenko and Porter}

\begin{document}

\title{The Gamma-ray Albedo of the Moon}

\author{Igor V. Moskalenko\altaffilmark{1}}
\affil{
   Hansen Experimental Physics Laboratory, 
   Stanford University, Stanford, CA 94305
\email{imos@stanford.edu}}
\altaffiltext{1}{Also Kavli Institute for Particle Astrophysics and Cosmology,
Stanford University, Stanford, CA 94309}

\and

\author{Troy A. Porter}
\affil{
  Santa Cruz Institute for Particle Physics,
  University of California, Santa Cruz, CA 95064
\email{tporter@scipp.ucsc.edu}}

\begin{abstract}

We use the \geant\ Monte Carlo framework to calculate the \gray{} 
albedo of the Moon due to interactions of cosmic ray (CR) 
nuclei with moon rock.
Our calculation of the albedo spectrum agrees with the EGRET data. 
We show that
the spectrum of \gray{s} from the Moon 
is very steep with an effective cutoff around 3--4 GeV (600 MeV
for the inner part of the Moon disk) and exhibits a narrow
pion-decay line at 67.5 MeV, perhaps unique in astrophysics. 
Apart from other astrophysical sources,
the albedo spectrum of the Moon is well understood, including
its absolute normalisation; this makes it a useful
``standard candle'' for \gray{} telescopes.
The steep albedo spectrum also provides a unique opportunity for energy 
calibration of \gray{} telescopes, such as the 
forthcoming Gamma Ray Large Area Space Telescope (GLAST).
Since the albedo flux depends on the incident CR spectrum which
changes over the solar cycle, it is possible
to monitor the CR spectrum using the albedo \gray{} flux.
Simultaneous measurements of CR
proton and helium spectra by the Payload for Antimatter-Matter 
Exploration and Light-nuclei Astrophysics (PAMELA), 
and observations of the albedo 
\gray{s} by the GLAST Large Area Telescope (LAT), can be used to test the model
predictions and will enable the LAT to monitor the
CR spectrum near the Earth beyond the lifetime of the PAMELA.

\end{abstract}


\section{Introduction}
Interactions of Galactic CR nuclei with the atmospheres of the Earth and 
the Sun produce albedo \gray{s} due to the
decay of secondary neutral pions and kaons 
\citep[e.g.,][]{Seckel1991,Orlando2007}.
Similarly, the Moon emits \gray{s} due to CR interactions
with its surface \citep{Morris1984,Thompson1997}.
However, contrary to the CR interaction with the gaseous
atmospheres of the Earth and the Sun, the Moon surface is solid, 
consisting of rock, making its albedo spectrum unique.

Due to the kinematics of the collision, the secondary particle cascade 
from CR particles hitting the Moon surface at small zenith angles
develops deep into the rock
making it difficult for \gray{s} to get out.
The spectrum of the albedo \gray{s} is thus necessarily soft 
as it is produced by a small fraction of low-energy splash particles
in the surface layer of the moon rock.
The high energy \gray{s} can be
produced by CR particles hitting the Moon surface with a more 
tangential trajectory.
However, since it is a solid target, only the very thin
limb contributes to the high energy emission.

The \gray{} albedo of the Moon has been calculated by \citet{Morris1984} 
using a Monte Carlo code 
for cascade development in the Earth's atmosphere that was modified 
for the Moon conditions.
However, the CR spectra used as input in the \citeauthor{Morris1984} 
calculation differ considerably
from recent measurements by AMS 
\citep{p_ams,he_ams} and BESS \citep{sanuki00} 
at both low and high energies. 
In particular, 
the CR proton spectrum used by \citeauthor{Morris1984} produces the 
correct intensity only around 10 GeV. 
Additionally, due to the lack of accelerator data and models a number 
of approximations and ad-hoc assumptions were required to calculate
the hadronic cascade development in the solid target of the Moon's surface.

The Moon has been detected by the EGRET as a point source 
with integral flux $F(>$$100\ {\rm MeV})=(4.7\pm0.7)\times10^{-7}$ cm$^{-2}$
s$^{-1}$ \citep{Thompson1997}, $\sim$24\% below the predictions by 
\citet{Morris1984} although the spectral shape agrees with the data. 
The observed spectrum is steep and yields only the upper limit 
$\sim$$5.7\times10^{-12}$ cm$^{-2}$ s$^{-1}$ above 1 GeV. 
At lower energies,
\gray{} spectroscopy data acquired by the Lunar Prospector 
have been used to map the elemental composition of the Moon surface 
\citep{Lawrence1998,Prettyman2006}.

In this paper, we calculate the \gray{} albedo from the Moon
using the \geant\ software framework and discuss the consequences of 
its measurement by the upcoming GLAST mission \citep{Michelson2007,Ritz2007}. 
Our preliminary results are presented in \citet{MP2007}.

\section{Monte Carlo simulations}\label{MC}

\begin{figure}
\centerline{
\includegraphics[width=2.5in]{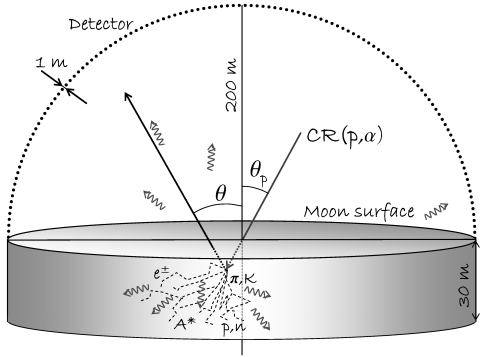}}
\caption{Beam/target/detector setup for simulating CR interactions in moon 
rock. 
The primary beam enters the moon rock target with incident polar 
angle $\theta_p$. 
Secondary \gray{s} are emitted with polar angle $\theta$. 
The detection volume surrounds the target.}
\label{fig1}
\end{figure}

\placefigure{fig1}

Since the first calculation of the Moon albedo \citep{Morris1984},
computers and Monte Carlo codes have advanced 
considerably, while more data on particle interactions 
and CR fluxes have become available.
The modern \geant\ toolkit\footnote{Available from http://cern.ch/geant4}
\citep{Agostinelli2003} 
is widely used across particle physics, 
medical science, and space physics to simulate the passage of particles 
through matter.
It provides extensive functionality including physics models, tracking, 
geometry, and visualisation.
A full range of electromagnetic and hadronic physics processes over a wide 
energy range are available including long-lived particles.
A large set of materials and elements are also included. 
Together with the extensive documentation and examples, it is straightforward 
to describe and simulate complicated detectors and interactions. 

In the present work, we use version 8.2.0 of the \geant\ 
toolkit.
Figure~\ref{fig1} illustrates our beam/target/detector setup for 
simulating CR interactions in the Moon.
The primary CR beam (protons, helium nuclei) is injected at different
incident angles into a moon rock target.
We take the composition of the moon rock to be 
45\% SiO$_2$, 22\% FeO, 11\% CaO, 10\% Al$_2$O$_3$, 
9\% MgO, and 3\% TiO$_2$ by weight, consistent with mare basalt 
meteorites and Apollo 12 and 15 basalts 
\citep{Lawrence1998,Anand2003,Prettyman2006}.
A thin hemispherical detector volume surrounding the target is used to 
record the secondary \gray{} angular and energy distributions in the simulation.

In \geant\ the specification of the {\it physics list} is of considerable
importance since it is where all the physics to be used in the simulation is
defined.
In the physics list, the user defines all necessary particles in the 
simulation, assigns physics processes to each, and sets appropriate range
cuts for secondary production.
In \geant\ there is a distinction between processes and models: 
processes take care of particle decays and interactions while for any 
process there may be one or more models and cross section parameterizations.
For most processes (electromagnetic, elastic, decay) there is only one model.
However, for hadronic processes and, in particular, inelastic processes, 
more than one model can be used over different energy ranges.
The preference for one model over another is not specified in the 
documentation but depends on the particular application.
Fortunately, to guide users of the toolkit, extensive examples are 
provided for use cases in different settings.

To construct our physics list, we examined the two examples in the 
\geant\ distribution that most closely resembled our simulation setup: 
the ``cosmicray\_charging'' and ``radioprotection'' examples in the 8.2.0 
release.
We included all relevant particles (leptons, photons, pions, kaons, 
heavy mesons, protons, antiprotons, neutrons, 
antineutrons, and resonances) and included the standard electromagnetic 
processes (ionisation, bremsstrahlung, multiple scattering, etc.) 
appropriate for each particle.
For hadronic processes, we included ``at rest'' processes, such as 
negative pion and kaon absorption, neutron and muon capture, 
antiproton and antineutron annihilation, and ``in flight'' processes such
as elastic and inelastic scattering (photo- and electro-nuclear and 
other secondary production), in-flight capture, and fission.
For all long-lived unstable particles the decay process was also included.

For inelastic processes 
\geant\ provides different types of hadronic shower
models, each of which has an intrinsic range of applicability.
Three types of hadronic shower models are available: data-driven models for 
high precision treatment of the transport of low energy neutrons, 
parameterised models based on a re-implementation of the GHEISHA package
of GEANT3.21, and theory-based models that cover a wide energy range.
Pre-packaged combinations of these models are provided with the \geant\
distribution.
However, for all of these inelastic scattering by ions is only 
treated up to 100 MeV. 
Cosmic-ray helium is a significant component of the CR flux 
interacting
in the Moon's surface, and we consider beam energies considerably higher
than this in our simulations. 
Furthermore, many of the pre-packaged lists are intended for use in 
applications such as high energy collider simulations where the treatment 
of low energy processes is less important.

Guided by the aforementioned examples and the documentation, we constructed
the hadronic physics list using the following models.
For energies up to 70 MeV, a pre-equilibrium decay and 
evaporation (PreCo) model for interactions of protons and neutrons was used.
Above this, a Bertini cascade (BERT) model between 60 MeV and 
4 GeV, a parameterised model (LEP) from 3 GeV to 15 GeV, and a 
quark gluon string (QGS) model from 12 GeV up to 100 GeV were employed.
For pions and kaons, the BERT model was used up to 4 GeV with 
the LEP and QGS models used over the same energy ranges as for protons
and neutrons.
Deuterons, tritons, and alpha particles use the low energy inelastic code 
up to 100 MeV, and a binary ion cascade model from 40 MeV to 40 GeV.
Each model is used in its applicable energy range 
(D. Wright, private communication).
In the energy ranges where there are overlaps,
the \geant\ code interpolates between the models.
The overlap ranges were chosen to ensure as smooth a transition between the 
models as possible.
These models compare well with data and other particle transport 
codes in standard accelerator and space environments 
\citep{Wellisch2003,Ersmark2004}\footnote{see also physics lists 
at http://geant4.slac.stanford.edu/
and http://www.particle.kth.se/desire/}.

The \gray{} yield $dY_\gamma(E_p,\cos\theta_p)/dE_\gamma d\cos\theta$
is calculated using the \geant\ beam/target setup with
a Monte Carlo method; here $E_p$ is the kinetic energy per nucleon of 
the incident particle, $\theta_p$ is the incident polar angle,
$E_\gamma$ is the energy of \gray{s}, and $\theta$ is the 
polar photon emission angle.

\begin{figure*}
\centerline{\includegraphics[width=3.5in]{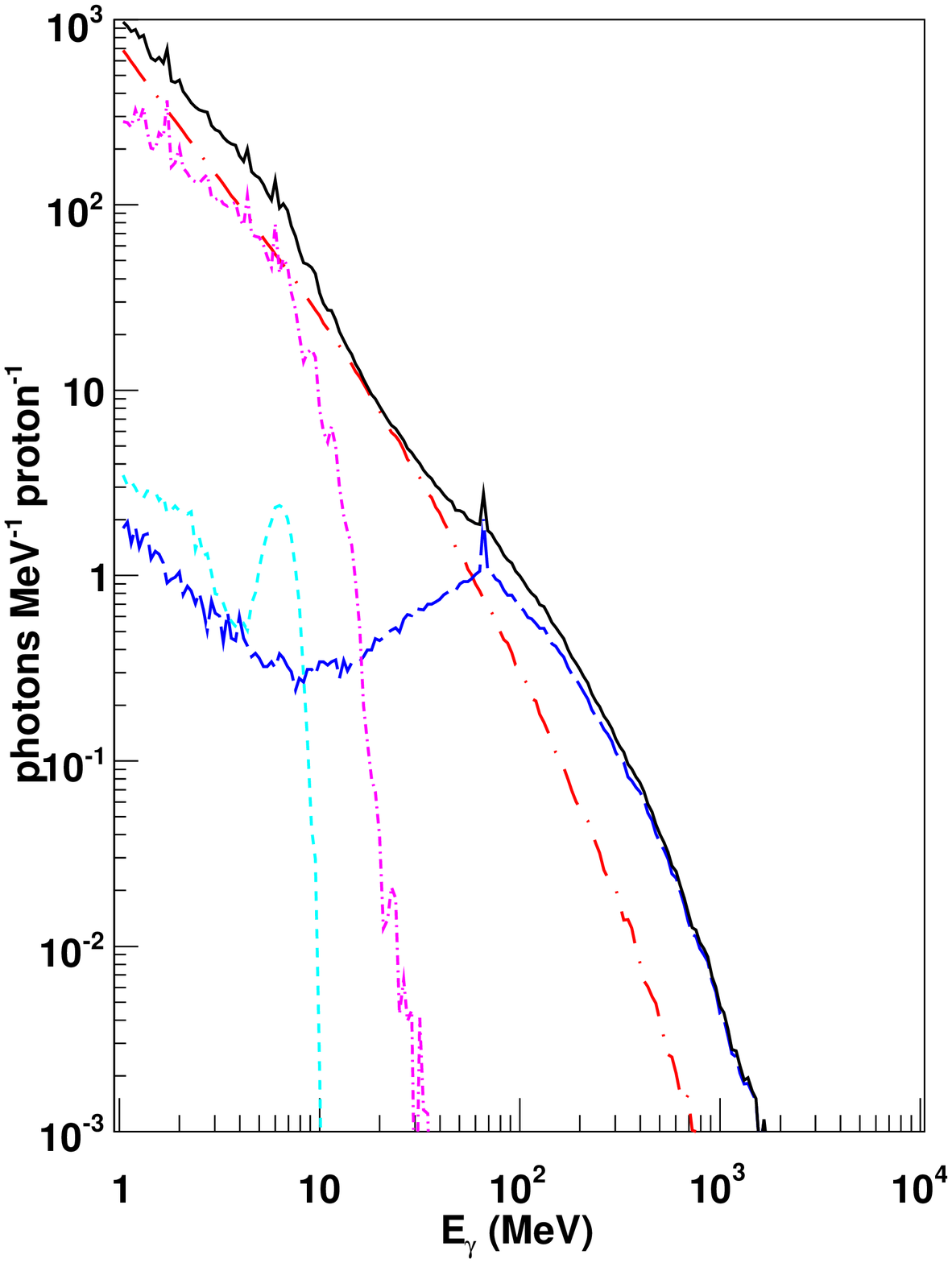} 
\includegraphics[width=3.5in]{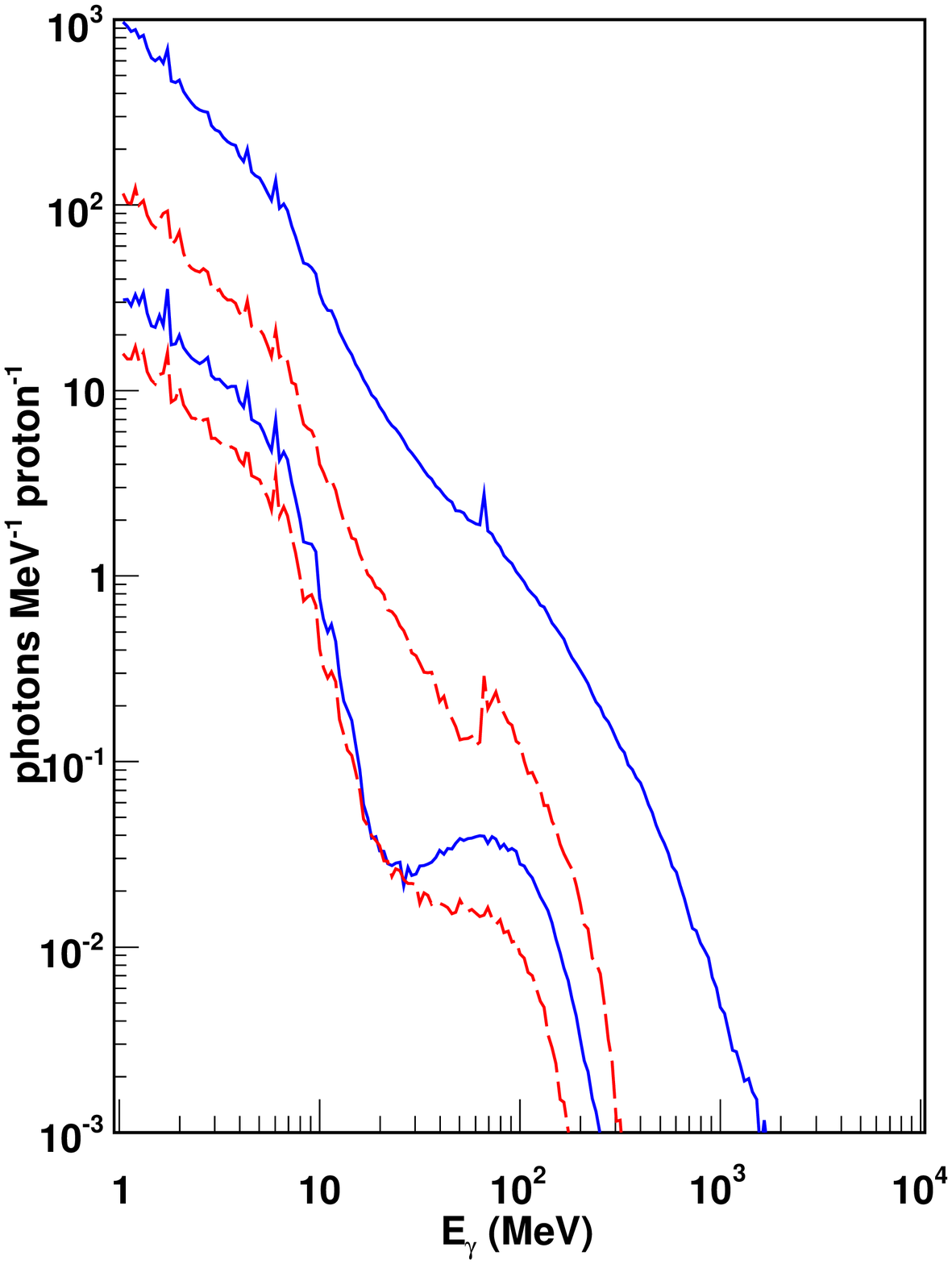}}
\caption{\gray{} yield per proton interaction integrated over
all emission angles from the Moon surface. 
Left panel: yield calculated for $E_p = 10000$ MeV 
and $\cos\theta_p = 0.1$ with components shown.
Line styles: blue-dashed, decay; red-long-dash-dot,
bremsstrahlung; magenta-short-dash-dot, non-elastic scattering and
de-excitation; cyan-dot, low energy neutron capture; black-solid, total.
For each component, the initial process responsible for the production of the
\gray\ secondary is given by the line style but the yield distribution 
includes also processes such as Compton scattering.
Right panel: yield calculated for two different energies and incident angles.
Line styles: blue-solid, $\cos\theta_p = 0.1$; 
red-dashed, $\cos\theta_p = 1$. 
Line-sets: lower, $E_p = 500$ MeV; upper, $E_p = 10000$ MeV.}
\label{fig2}
\end{figure*}

Figure~\ref{fig2} (left) 
shows the total secondary \gray{} yield and the contribution
by different processes integrated over all 
emission angles outward from the Moon's surface for protons with 
$E_p = 10000$ MeV and incident angle $\cos\theta_p = 0.1$. 
The \gray\ emission is produced in a number of processes:
pion and kaon decay, secondary electron and positron bremsstrahlung, 
and so forth. 
A considerable flux of \gray{s} is produced in nuclear 
reactions such as neutron capture and nonelastic scattering
\citep{Lawrence1998,Prettyman2006};
the features below $\sim$10 MeV are due to nuclear de-excitation lines,
where the most prominent contribution comes from oxygen nuclei.
The development of the cascade in the rock causes modification of the yield
spectra from the usual spectral distributions.
For example, the $\pi^0$-decay spectrum would normally be symmetric 
about 67.5 MeV, instead it has a significant low-energy extension
due to Compton scattering of the \gray{s} in the rock.

\placefigure{fig2}

Figure~\ref{fig2} (right) shows the secondary \gray{} yields integrated over 
all emission angles outward from the Moon surface for
protons with $E_p = 500$ MeV and 10000 MeV at incident 
angles $\cos\theta_p = 0.1$ and 1, respectively.
The shapes of the yield curves for different incident angles are very similar
to each other for the case of low energy protons where the secondary particles
(pions, kaons, neutrons, etc.) are produced nearly at rest. 
In the high energy case,
the secondary distribution for protons incident near zenith has a 
cutoff above $\sim$500 MeV.
Further away from zenith, higher yields of secondary \gray{s} are 
produced while the spectrum of \gray{s} becomes
progressively harder.
This is a result of the cascade developing mostly in the forward direction:
for near zenith primaries, most high energy secondary \gray{s} will be 
absorbed in the target, while a small fraction of 
splash albedo particles, mostly low energy ones, produce the 
soft spectrum \gray{s}; further from zenith, the high energy secondary
\gray{s} will shower out of the Moon surface.

The clear narrow line at 67.5 MeV in Figure~\ref{fig2} (left and right panels) 
is due to neutral pions decaying near rest.
To ensure that it is not an artefact of the simulation, we made an
extensive analysis of the hadronic interaction models employed
in \geant\ and various geometries of the beam/target/detector 
configuration. 
Independently of the hadronic interaction model used,
the line is always present, but its intensity may vary slightly.
However, dependent on the beam/target/detector configuration 
the line may appear or disappear. 
The most feasible explanation of this effect is the following.
The line is always present since there is always some small fraction of pions
decaying at rest, but in the most common configurations, such as
thin target, gaseous target, and/or detector position in the forward
hemisphere, the line is hidden under the strong continuum background.
In our case of a solid thick target and the detector location 
in the backward hemisphere, a considerable part of the continuum 
background is cut off and the line becomes visible.

There are several reasons why the continuum background is cut off.
The interaction length of a GeV proton in oxygen 
(the most abundant element in the moon rock) is $\sim$80 g cm$^{-2}$,
which corresponds to $\sim$25 cm for the moon rock 
(density $\sim$3.3 g cm$^{-3}$).
As the cascade develops, energetic particles
penetrate deep into the rock, too deep for \gray{s} to come out.
The \gray{} albedo is produced by relatively low-energy
particles at the depth of about one radiation length, $\sim$34 g cm$^{-2}$,
which corresponds to $\sim$10 cm for the moon rock. 
The pions decaying
in the upper layer of the moon rock are necessarily
low-energy (splash) particles 
while the high-energy part of the cascade producing the 
continuum spectrum is cut off because of the kinematics of the
interaction and the forward cascade development.
Since many photons are coming from a large depth, this
explains the large contribution of the Compton scattered \gray{s}
below $\sim$60 MeV (Figure~\ref{fig2}).
For lower energy protons, this also explains why the 67.5 MeV line is not prominent, 
since the development of the cascade then occurs relatively close to the 
surface producing a continuum that essentially washes out any line feature.

\section{Calculations}
The CR spectrum above the geomagnetic cutoff 
near Earth (at 1 AU) can be directly measured by 
balloon-borne instruments or spacecraft. 
However, it has been done
during short flights at different phases of solar activity.
To calculate the Moon albedo at an arbitrary modulation level, we use
the local interstellar (LIS) spectra of CR protons and helium as fitted to the 
numerical
results of the GALPROP propagation model 
(reacceleration and plain diffusion models, Table 1 in \citet{Ptuskin2006});
the CR particle flux at an arbitrary phase of solar activity at 1 AU
can then be estimated using the force-field approximation \citep{Gleeson1968}:

\begin{equation}
\frac{dJ_p(E_k)}{dE_k}=\frac{dJ_p^\infty(E_k+\Phi Z/A)}{dE_k}\frac{E^2-M^2}{(E+\Phi Z/A)^2-M^2},
\end{equation}
where $dJ_p^\infty/dE_k$ is the LIS spectrum of CR species, $E_k$ is the
kinetic energy per nucleon, $E$ is the total energy per nucleon,
$\Phi$ is the modulation potential, $Z$ and $A$ are the nucleus charge
and atomic number correspondingly, and $M$ is the nucleon mass.  

\begin{figure}
\centerline{
\includegraphics[width=3.5in]{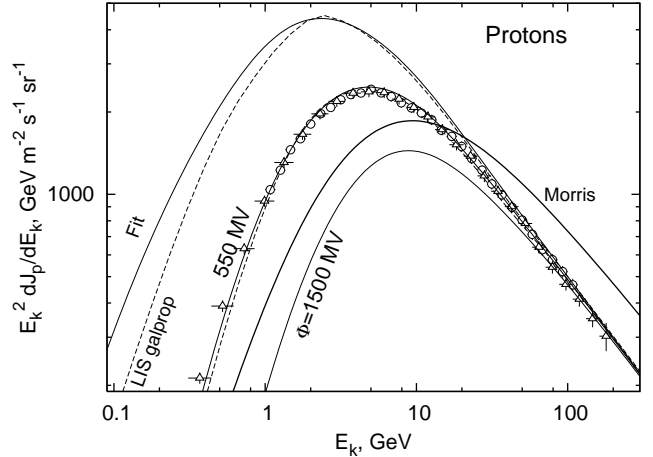}}
\caption{The CR proton spectrum. The dashed lines are the LIS spectrum
(upper) and modulated one (lower, $\Phi = 550$ MV) as calculated by GALPROP 
in the plain diffusion model 44\_999726 \citep{Ptuskin2006}. The thin
solid lines are the fit to the LIS spectrum (upper) and modulated spectra (middle,
$\Phi = 550$ MV; lower $\Phi = 1500$ MV). The thick solid line is the CR proton
spectrum used by \citet{Morris1984}. Data: circles --- BESS \citep{sanuki00}, triangles
--- AMS \citep{p_ams}.}
\label{fig3}
\end{figure}

To fit the LIS CR spectra we choose a function of the form:
\begin{equation}
\frac{dJ^\infty}{dE_k}=J_0 \sum_{i=1}^3 a_i (E_k+b_i)^{-c_i},
\label{fit}
\end{equation}
where the flux units are m$^{-2}$ s$^{-1}$ sr$^{-1}$ (GeV/nucleon)$^{-1}$
and the parameter values are given in Table~\ref{Table1}.
The latter are not unique and other sets could produce similar
quality fits, but this does not affect the final results.
Figure \ref{fig3} shows the LIS CR proton spectrum and modulated one 
compared with the
BESS \citep{sanuki00} and AMS \citep{p_ams} data taken during
the period of moderate solar activity. 
The parameters given in Table~\ref{Table1} yield a LIS proton
spectrum that is somewhat \emph{higher} than the GALPROP LIS below $\sim$2
GeV \citep[plain diffusion model 44\_999726,][]{Ptuskin2006}.
However, the fit to the AMS and BESS data after modulation 
($\Phi = 550$ MV) is good. The spectrum labelled $\Phi = 1500$ MV
corresponds to the period of high solar activity. 
The proton spectrum used
by \citet{Morris1984} is shown by the thick solid line. 
The disagreement
with the modern CR proton data is considerable.
For the helium spectrum, the parameters given in Table~\ref{Table1} provide
a good fit to the GALPROP LIS, and agree with the AMS \citep{he_ams} 
data after modulation ($\Phi = 550$ MV); at high energies the fit also 
passes through JACEE \citep{he_jacee} and Sokol \citep{he_sokol} data.

\placetable{Table1}

\begin{deluxetable*}{lcccccccccc}
\tablecolumns{11}
\tablewidth{0pc}
\tabletypesize{\footnotesize}
\tablecaption{Fits to local interstellar CR spectra \label{Table1}}
\tablehead{
\colhead{particle} &
\colhead{$J_0$} &
\colhead{$a_1$} & \colhead{$b_1$} & \colhead{$c_1$} &
\colhead{$a_2$} & \colhead{$b_2$} & \colhead{$c_2$} &
\colhead{$a_3$} & \colhead{$b_3$} & \colhead{$c_3$} 
}
\startdata
proton & $1.6 \times 10^4$ & 1 & 0.458 & 2.75 & --3.567 & 0.936 & 4.90 & $4.777\times10^5$ & 14.4 & 6.88 \\
helium & $1.6\times10^3$ & 1 & 1.116 & 3.75 & 2.611 & 4.325 & 3.611 & 0.219 & 0.923 & 2.58
\enddata
\tablecomments{Intensity units: m$^{-2}$ s$^{-1}$ sr$^{-1}$ (GeV/nucleon)$^{-1}$.}
\end{deluxetable*}

\placefigure{fig3}

The \gray{} albedo flux at the Earth is calculated as:
\begin{equation}
\frac{dF_\gamma}{dE_\gamma d\cos\psi}
=\frac1{\cos\theta}\int dE_k\, \frac{dJ_p}{dE_k}
\frac{d\tilde{Y}_\gamma(E_k,\cos\psi)}{dE_\gamma d\cos\theta},
\end{equation}
where $\psi$ is the angular distance from the geometrical center of the
Moon disk as seen from the Earth, $1/\cos\theta$ is the Jacobian,
\begin{equation}
\cos\theta=\sqrt{1-\frac{R^2}{r^2}\sin^2\psi},
\end{equation}
\begin{equation}
\frac{d\tilde{Y}_\gamma(E_k)}{dE_\gamma d\cos\theta}=2\pi \int_0^1 d(\cos\theta_p)\, 
\cos\theta_p \frac{dY_\gamma(E_k,\cos\theta_p)}{dE_\gamma d\cos\theta}, 
\end{equation}
$R=1738.2$ km is the Moon radius, and $r=384401$ km is the distance to the Moon.

\begin{figure}
\centerline{
\includegraphics[width=3.5in]{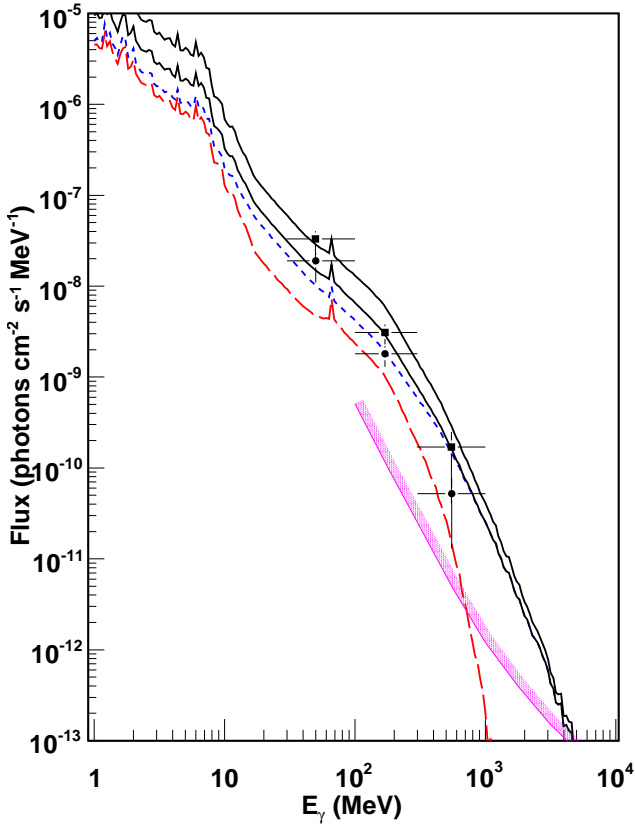}}
\caption{Calculated \gray{} albedo spectrum of the Moon.
Line-styles: black-solid, total; blue-dotted, limb -- outer $5'$; 
red-dashed, center -- inner $20'$.
Upper solid line: $\Phi = 500$ MV; lower solid line: $\Phi = 1500$ MV.
Limb and center components are only shown for $\Phi = 1500$ MV.
Data points from the EGRET \citep{Thompson1997} 
with upper and lower symbols corresponding 
to periods of lower and higher solar activity, respectively.
The differential 1 year sensitivity of the 
LAT is shown as the shaded region.}
\label{fig4}
\end{figure}

Figure \ref{fig4} shows the calculated total \gray{} albedo spectrum 
for CR protons and helium compared to the EGRET data for periods of lower 
(upper solid: $\Phi = 500$ MV) and higher 
(lower solid: $\Phi = 1500$ MV) solar activity. Taking into 
account that the exact CR spectra during the EGRET observations
are unknown, the agreement with the data is remarkable.
The broken lines show the spectra
from the limb (outer $5'$) and the central part of the 
disk ($20'$ across) for the case of higher solar activity.
As expected, the spectra from the limb and the
central part are similar at lower energies ($<$10 MeV); at high
energies the central part exhibits a softer spectrum so that virtually
all photons above $\sim$600 MeV are emitted by the limb.

Interestingly, the pion-decay line is still significant even when the
\gray{} yields are integrated over the spectrum of CR,
although, its intensity is small relative to the \emph{total} albedo flux.
The line perhaps always exists due to the splash albedo pions,
but in case of a thin and/or gaseous target (usual in astrophysics)
it is indistinguishable from the background \gray{s}
(see Section~\ref{MC} for more discussion).

\placefigure{fig4}

\section{Discussion}
The GLAST LAT 
is scheduled for launch by NASA in early 2008. 
It will have 
superior angular resolution and effective area, and its field of view (FoV)
will far exceed that of its predecessor, the EGRET \citep{McEnery2004}.
The LAT will scan the sky continuously providing complete sky coverage
every two orbits (approximately 3 hr). 
The on-axis effective area of the LAT increases from $\sim$3000 cm$^2$ at 
100 MeV to $\sim$8500--9000 cm$^2$ at 1 GeV and 
higher\footnote{see http://glast.stanford.edu}.
In this case, the point spread function (PSF) of the instrument has a 
68\% containment radius $\sim$$4.3^\circ$ at 100 MeV reducing 
dramatically at higher energies: $\sim$$0.8^\circ$ at 1 GeV,
$\sim$$0.5^\circ$ at 2 GeV, and $\sim$$0.2^\circ$ at 10 GeV.
Using only events from the front section of the LAT, the 
PSF improves to a 68\% containment radius of $\sim$$3^\circ$ at 100 MeV,
$\sim$$0.5^\circ$ at 1 GeV, 
$\sim$$0.25^\circ$ at 2 GeV, and $\sim$$0.1^\circ$ at 10 GeV, but the on-axis
effective area is essentially halved.

About 20\% of the time the Moon 
will be in the FoV at different viewing angles.
The photon flux expected from the Moon for $\Phi = 500$ MV 
above 100 MeV, 1 GeV, and 4 GeV is 
$\sim$$5\times10^{-7}$, $\sim$$2\times10^{-8}$, and $\la$$1.6\times10^{-10}$ 
photons cm$^{-2}$ s$^{-1}$, respectively.
For $\Phi = 1500$ MV, the expected flux is reduced by a factor of 2 at 100 MeV
only.
With these fluxes and the above values for the effective area, and allowing 
for an additional factor 2 reduction to take into account time off-axis, 
instrumental deadtime, and South Atlantic Anomaly traversals, 
we estimate the LAT will collect $\sim$$5\times10^3$, $\sim$$6\times10^2$, 
and $\la$5 photons, respectively, in the above energy ranges in one year.
These numbers are reduced by a factor of two if only events from the front 
section of the LAT are used. 
Interestingly, the albedo flux at low energies 
is high enough that the impact of the broader PSF is significantly reduced.

Measuring \gray{s} from the Moon presents several interesting 
possibilities for the LAT.
Apart from other astrophysical sources,
the albedo spectrum of the Moon is well understood, including
its absolute normalisation, 
while the Moon itself is a ``moving target'' passing through high Galactic 
latitudes and the Galactic centre region. 
This makes it a useful
``standard candle'' for the GLAST LAT at energies where the PSF
is comparable to the Moon size $0.5^\circ$, i.e., at $\sim$1 GeV 
and higher. 
At these energies the albedo flux is essentially 
independent of the solar modulation.
At lower energies the \gray{} flux depends on the level of the solar 
modulation and thus can be used to infer the incident CR spectrum.
A simultaneous presence of the PAMELA
on-orbit capable of measuring protons and light nuclei with
high precision \citep{Picozza2006} provides a necessary input for accurate
prediction of the albedo flux and a possible independent 
calibration of the GLAST LAT.
An additional bonus of such a calibration is the possibility to use
the GLAST observations of the Moon to monitor the CR spectra
near the Earth beyond the projected lifetime of the PAMELA (currently
3 years).

The line feature at 67.5 MeV  from $\pi^0$-decay 
produced by CR particles in the solid rock target is interesting.
There is no other astrophysical object predicted to
produce such a narrow line and there is no other line expected
except, perhaps, from dark matter annihilation.
The lower energy limit of the LAT instrument
is below 20 MeV while the energy resolution is $\sim$15\% at 100 MeV, and 
improves at higher energies.
With a suitable event selection it may be possible to observe the line;
if so, it will provide a possibility of in-orbit energy calibration.
Another possibility for energy calibration at higher energy is provided
by the steep albedo spectrum above 100 MeV:
a small error in the energy determination will result in a large
error in the intensity.

Because of the steep albedo spectrum,
very few photons will be detected above $\sim$3--4 GeV.
The central part of the Moon has an even steeper 
spectrum with an effective cutoff at $\sim$600 MeV.
Above a few GeV, the Moon 
presents almost a black spot on the \gray{} sky 
providing an opportunity to screen out a piece of the sky. 
Additionally, the expected irreducible background rate\footnote{\gray{s}
produced in the FoV by CR interactions in the inert material 
surrounding the LAT.} at these energies within the solid angle subtended
by the Moon is $\sim$0.1 photons per year 
(S. Digel, private communication).
The absence of any significant excess over the predicted rate will be a 
useful cross check on the expected irreducible background.

\section{Conclusions}

The \gray{} albedo of the Moon makes
it a unique calibration target for \gray{} telescopes.
Apart from other astrophysical sources 
the albedo spectrum is well understood, including
its absolute normalisation.
The albedo of the Moon is dim, especially
its central part at high energies.
Its lower energy part exhibits a narrow pion-decay line at 67.5 MeV, 
perhaps unique in astrophysics and never before observed,
while its continuum intensity depends on the phase of the
solar cycle and allows one to monitor the ambient spectrum of CR
particles.
The GLAST LAT instrument is well suited for such observations.

\acknowledgments
We thank Bill Atwood, Seth Digel, Robert Johnson, 
and Dennis Wright for many fruitful discussions and 
the anonymous referee for useful suggestions.
I.\ V.\ M.\ acknowledges partial support from NASA
Astronomy and Physics Research and Analysis Program (APRA) grant.
T.\ A.\ P.\ acknowledges partial support from the US Department of Energy.

\end{document}